\def\be {\begin{equation}}
\def\ee {\end{equation}}

\def\R{{\bf R}}

\documentstyle[preprint,eqsecnum,aps]{revtex}
\begin{document}
\tighten
\preprint{UCSBTH-96-03, gr-qc/9602036}
\title{Interpolating Between Topologies: Casimir Energies}
\author{Donald Marolf}
\address{Physics Department, The University of California,
Santa Barbara, California 93106} \date{February, 1996}
\maketitle

\begin{abstract}
A set of models is considered which, in a certain sense, interpolates
between 1+1 free quantum field theories on topologically distinct
backgrounds.  The intermediate models may be termed free quantum field
theories, though they are certainly not local.  Their ground
state energies are computed and shown to be finite.  The possible
relevance to changing spacetime topologies is discussed.

\end{abstract}

\vfil
\eject
\section{Introduction}
\label{intro}

The suggestion that the topology of spacetime is not 
fixed \cite{JW}, but may change in the course of (quantum)
evolution has attracted much interest and has been addressed from
many points of view.  Perhaps the earliest concrete work on the
subject was that of Anderson and DeWitt \cite{AD} (followed
by \cite{MD1}) who calculated
the effect of a topology changing background on a propagating
scalar quantum field.  Finding an infinite production of energy, they
concluded that topology change was heavily suppressed.

However, in their analysis, the background changed instantly from one
topology to the other.  It is not surprising that a discontinuous
process in quantum field theory
produces an infinite amount of energy; indeed, this is what
one would expect from the naive estimate $E \approx \hbar/ T$ for 
$T=0$. 

A number of attempts may be made to smooth out this transition.
One idea is to use a region of
Riemannian signature spacetime to connect the initial and final 
topologically different regions of Lorentzian spacetime.
Some steps in this direction were accomplished in \cite{MD2,MD3,MD4,LS}.
Another approach is to consider 2+1 gravity which, when
properly formulated, \cite{Witten0} contains {\it classical} solutions which
change topology; both in vacuum \cite{gary} and with matter
fields \cite{matter}.  However, it is not clear how to put quantum
fields on such backgrounds and 
Euclidean arguments in 2+1 quantum gravity have been used both for 
\cite{Witten,AH} and against \cite{Carlip} topology change.  As a
final remark, the topology
of spacetime appears to fluctuate in string theory \cite{string}. 

Our goal here is to address this matter from a new perspective by
considering a class of theories that, in a certain sense, 
interpolate between field
theories on different manifolds.  Our systems will clearly not
be local, but they will be `Lorentzian' quantum field theories possessing a
Hilbert space of states, unitary dynamics, and finite ground state
energies.  Each theory will be characterized by some matrix
$M \in U(2)$, with the local free scalar field theories  on
$S^1 \times \R$ and on $(S^1 \times \R) \cup (S^1 \times \R)$ being
represented by particular values of $M$.  Since the set $U(2)$ is
connected, our class of quantum fields can be said to interpolate
between these two topologies.  The interpolating theories are
described below.

\section{The Interpolating Theories}

The original Anderson-DeWitt calculation is
reminiscent of turning on a mirror in 1+1 spacetime.  In their case, the mirror
was turned on quickly, that is, `all at once.'  We wish to study the
intermediate cases between the initial and final topologies.  As a
result, our models will be the topology change analogues of 
partially silvered mirrors.

The models we consider contain a single free scalar field on a 
1+1 manifold whose spatial `hypersurfaces' consist of two 
disconnected line segments.  As such, our field lives on the manifold
${\cal M} = (I \cup I) \times \R$ with boundary $\R \cup \R \cup \R \cup \R$.
The various components of the boundary will be thought of as 
the worldlines of `gates' into which flux from the field is allowed
both to enter and to emerge.  That is, the dynamics of our system
will be defined so that flux entering one of the gates immediately
reemerges from another.  The details of these connections will
be given by a `routing' matrix $M$ (described below),
with different routing matrices
corresponding to different kinds of gates.
We will take our gates to be independent of both time and frequency.

\

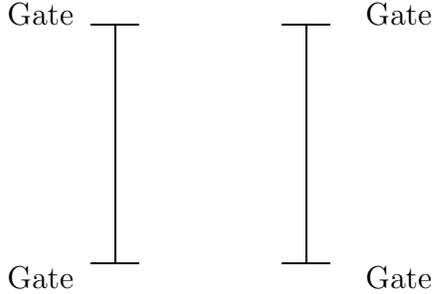
\setlength{\unitlength}{0.01250000in}%
\begingroup\makeatletter\ifx\SetFigFont\undefined
\def\x#1#2#3#4#5#6#7\relax{\def\x{#1#2#3#4#5#6}}%
\expandafter\x\fmtname xxxxxx\relax \def\y{splain}%
\ifx\x\y   
\gdef\SetFigFont#1#2#3{%
  \ifnum #1<17\tiny\else \ifnum #1<20\small\else
  \ifnum #1<24\normalsize\else \ifnum #1<29\large\else
  \ifnum #1<34\Large\else \ifnum #1<41\LARGE\else
     \huge\fi\fi\fi\fi\fi\fi
  \csname #3\endcsname}%
\else
\gdef\SetFigFont#1#2#3{\begingroup
  \count@#1\relax \ifnum 25<\count@\count@25\fi
  \def\x{\endgroup\@setsize\SetFigFont{#2pt}}%
  \expandafter\x
    \csname \romannumeral\the\count@ pt\expandafter\endcsname
    \csname @\romannumeral\the\count@ pt\endcsname
  \csname #3\endcsname}%
\fi
\fi\endgroup
\centerline{
\begin{picture}(150,119)(75,569)
\thinlines
\put(210,580){\line(-1, 0){ 20}}
\put(210,680){\line(-1, 0){ 20}}
\put(110,680){\line( 1, 0){ 20}}
\put(130,580){\line(-1, 0){ 20}}
\put(200,580){\line( 0, 1){100}}
\put(200,680){\line( 0, 1){  0}}
\put(120,680){\line( 0,-1){100}}
\put( 75,680){\makebox(0,0)[lb]{\smash{\SetFigFont{12}{14.4}{rm}Gate}}}
\put( 75,570){\makebox(0,0)[lb]{\smash{\SetFigFont{12}{14.4}{rm}Gate}}}
\put(225,570){\makebox(0,0)[lb]{\smash{\SetFigFont{12}{14.4}{rm}Gate}}}
\put(225,680){\makebox(0,0)[lb]{\smash{\SetFigFont{12}{14.4}{rm}Gate}}}
\end{picture}
}

For the appropriate routing matrices our system is equivalent
to a (local) 1+1 field theory on some manifold.  For example, if flux entering a
gate reemerges from the {\it same} gate, perhaps with a phase shift, 
then we have a field theory on $(I \cup I) \times \R$ with
boundary conditions of the usual type (Dirichlet, Neumann, or a 
combination).  By connecting the gates in different ways, we may
also construct field theories on closed manifolds.

Of course, to define a satisfactory quantum field theory requires
more than just a well-defined dynamics for the field.  We also require
a notion of unitarity which, in the context of free field theory, is
usually taken to be conservation of the Klein-Gordon flux.  Since
a general gated theory will be non-local, conservation of the
Klein-Gordon current may become a bit subtle at the gates.  
However,  for appropriately constructed gates, 
the Klein-Gordon norm will be conserved when evaluated on any
hypersurface that respects the gates' notion of simultaneity, as we
shall explain below.  The key point of our study is that
the class of gates for which the Klein-Gordon norm is conserved
is a several parameter family which interpolates between different
spacetime topologies.  Furthermore, these field theories will be seen
to possess ground states whose finite energy is a continuous function 
of the parameters.  Thus we will see that field theories on different
topologies can in fact be connected, in what seems to be a continuous manner. 

We will assume that the line segments are oriented towards the top
of the page and that the gates respect this orientation.  That is, 
we assume that flux which enters one of the top gates must
emerge from the {\it bottom} gates, and vice versa.  The more
general case in which the upward and downward moving parts of the field
become mixed by the gates can also be studied and leads to a larger
class of field theories.  Such theories behave much the same
as the orientation-preserving cases but are significantly more
complicated. As a result, we will not consider them here.

Our first task will be to consider the corresponding classical field
theories and find conditions under which the Klein-Gordon flux
is conserved.  It is convenient to give both line segments a coordinate
$x$ proportional to proper distance and
running from $0$ at the bottom to $\pi$ at the top; thus, both
segments are the same length.  Similarly, we use a coordinate
$t$ on $\R$ proportional to proper time. We consider
a free 1+1 scalar field $\phi_I$ in the interior of the first
segment and a field $\phi_{II}$ on the interior of the second segment.
These fields may be
separated into their upward and downward moving parts $\phi^+_{I,II} (x-t)$
and $\phi^-_{I,II}(x+t)$ as usual.  Having assumed that our gates
respect this decomposition, the details are
described by the boundary conditions they impose: 

\be
\label{+gate}
\phi^+(0,t) \equiv 
\left[  {{\phi_I^+ (x=0,t) } \atop {\phi_{II}^+ (x=0,t)}} \right]
= M_+ \left[ {{\phi_I^+ (x=\pi,t) } \atop {\phi_{II}^+ (x=\pi,t)}} \right]
\ee
and
\be
\label{-gate}
\phi^-(\pi,t) \equiv 
\left[  {{\phi_I^- (x=\pi,t) } \atop {\phi_{II}^- (x=\pi,t)}} \right]
= M_- \left[ {{\phi_I^- (x=0,t) } \atop {\phi_{II}^- (x=0,t)}} \right],
\ee
where this matrix notation will simplify a number of expressions below.
Note that each gate has an inherent notion of simultaneity in that
it connects some set of spacetime events.  We consider only
the case in which these notions
are all consistent in the sense that there is 
some Lorentz frame in which the gates connect only events with the same
value of time $t$.  We use coordinates 
adapted to this frame and
the matrices $M_\pm$ are taken to be time independent.

Let us first consider the
case where the downward-moving component vanishes.  Equating the
flux leaving the top gates with the flux entering the bottom 
gates leads to the condition
\be
0 ={\rm Im} \left\{
 [\phi^+(\pi,t)]^\dagger \ [ I - M_+^\dagger M_+ ] \
{\partial}_t \phi(\pi,t)
\right\}
\ee 
where ${\rm Im}$ denotes the imaginary part.
Since $\phi^+(\pi,t)$ and $\partial_t \phi^+(\pi,t)$
can be chosen independently at any time $t$, upward-moving flux
is conserved if and only if $M_+$ is unitary.  Similarly, conservation
of downward moving flux requires $M_-$ to be unitary.  Considering
conservation of the Klein-Gordon inner product $(\phi_+,\phi_-)$
of an upward
moving field $\phi_+$ with a downward moving field $\phi_-$
leads to the condition 
\be 
0 = \partial_t \left\{ [\phi_+(\pi,t)]^\dagger \  ( I - M^\dagger_+
M^{-1} ) \  \phi_-(\pi,t) \right\}
\ee
so that $M_+ = M^{-1}_-$.  These conditions are sufficient to guarantee
conservation of an arbitrary Klein-Gordon inner product.  Thus, 
any set of gates for which $M_+ = M^{-1}_-$ and for which $M_+$
is unitary may be considered to define a free (gated) quantum field theory.
While in general such theories are not `local' in any way, particular
examples include the case
\be 
\label{1}
M^{S^1}_+ = \left[ {0 \atop 1} {1 \atop 0} \right]
\ee
which is equivalent to
field theory on $S^1 \times \R$, the case
\be
\label{2}
M^{S^1 \cup S^1}_+ = \left[ {1 \atop 0} {0 \atop 1} \right]
\ee
which describes
field theory on two (smaller) copies of $S^1 \times \R$, the 
case
\be
\label{3} M^{S^1 \cup \tilde S^1}_+ = \left[ {-1 \atop 0} {0 \atop 1} \right]
\ee
which yields field theory on two copies of $S^1 \times \R$ with one 
circle having antiperiodic boundary conditions, 
as well as examples equivalent to other local theories.  
Since the unitary matrices are a (path) connected
set, any pair of these systems is linked by a continuous path through
our space of nonlocal theories.

In order to show that our field theories are `reasonable',
we now compute their ground state energies and show that 
they are finite and depend continuously on the routing matrix $M_+$.
Actually, we will not compute the ground state energy of all such
theories, but we will perform the computation for a large enough set
to form a continuous path between field theories on topologically
different manifolds.  

The reason for our restriction is that the dynamics of a general
theory is quite complicated; a generic gated theory does not
have periodic solutions of the form $e^{- i \omega (x \pm t)}$ for
{\it any} frequency omega.  However, such solutions do exist
when the gates are appropriately `tuned.'
To see this, note
that \ref{+gate} may be used to express the field $\phi^+_{II}(0,t)$
at the origin of the second segment in terms of the fields at
$x=0,\pi$ on the first segment.  Incrementing $t$ by $\pi$
expresses $\phi_{II}^+(\pi,t)$ in terms of $\phi_I^+(0,t + \pi)$ and
$\phi_I^+(0,t + 2 \pi)$.  Again using the top line of \ref{+gate} leads
us to the relation
\be
\label{per}
\left[ {{\phi^+_I(0,t)} \atop {\phi^+_I(\pi,t)}} \right]
= \left[ {{-\det M_+} \atop 0} \ {{{\rm Tr}M_+} \atop 1} 
\right]
\left[  {{\phi^+_I(2\pi,t)} \atop {\phi^+_I(\pi,t)}} 
\right].  
\ee
Thus, whenever ${\rm Tr} M_+$ vanishes $\phi^+_I(x,t)$ is periodic
up to a phase under $t \rightarrow t + 2 \pi$.  A corresponding
result for the downward-moving component follows by replacing
$M_+$ with $M_- = M^\dagger_+$.  The traceless unitary matrices are
again a connected set which, in particular (see \ref{1}, \ref{3}), 
interpolate between the free field on $S^1 \times \R$ and
the free field on two copies of $S^1 \times \R$ with one copy
having periodic boundary conditions and the other antiperiodic\footnote{
The reader may be surprised that the theory \ref{2} on two {\it
periodic} copies of $S^1 \times \R$ does not satisfy ${\rm Tr}M_+ = 0$.  
This occurs because ${\rm Tr}M_+ = 0$ is only a sufficient and not a necessary
condition for periodicity.  Nevertheless, it captures the generic
periodic case.}.  This set is thus large enough for our purposes.

It is straightforward to calculate the ground state
energy for each theory of this kind.  Note that
$(-\det M)$  is some phase $e^{i \theta}$ $(0 \leq \theta < 2 \pi )$.  
Suppose now that
each segment is of proper length $L$.  From \ref{per}, we see that 
the positive frequency
spectrum of the upward moving modes is $\{ {{ 2 n \pi + \theta}
\over {2 L}} : n \ge 0 \}$ while that of the downward moving
modes is $\{ {{2(n + 1)\pi  - \theta} \over {2 L}} : n \ge 0 \}$.
The ground state energy is then given by the 
expression
\begin{eqnarray}
\label{gs}
\lim_{\epsilon \rightarrow 0} & {\rm Re} &
\sum_{n \ge 0} \left[ e^{i \epsilon {{\theta}
\over {2L}}}  e^{i \left( {{\pi n} \over L} \right) \epsilon}
\left({{ 2 \pi n + \theta} \over { 4 L}} \right)  
+ e^{-i \epsilon {{ -\theta} \over {2L}}} 
e^{i \left( {{\pi (n+1)} \over L} \right) \epsilon} 
\left(  {{ 2 \pi (n+1) -\theta} \over {4 L}} \right)
+ {{2 L} \over {2 \pi \epsilon^2}} \right] \cr
& = & -
{\pi \over {12 L}} \left( {3 \over {2 \pi^2}} (\theta - \pi)^2 - {1
\over 2} \right)
\end{eqnarray}
where ${\rm Re}$ denotes the real part.  Note that \ref{gs}
is a finite and continuous function of $M_+$.

In particular, the cases described by \ref{1} and \ref{3} can be 
connected by a path along which every routing matrix satisfies
$(\det{M_+} = - 1)$.  As a result, the ground state energy is 
constant along this path.  This is consistent with known
expressions for the ground state energies of periodic and antiperiodic
fields on $S^1 \times \R$ \cite{Cas} and for taking the ground state
energy of a field on a disconnected manifold to be the sum of the
corresponding energies for its connected components. 
This consistency
with additivity provides a certain confirmation that the Minkowski vacuum, 
relative to which \ref{gs} is computed, is the correct zero of energy.

If we could consider a field theory in which the routing matrix
was not constant but instead was slowly varied as a function of
time, we might construct a system that smoothly evolves from
a field theory on $S^1 \times \R$ to a field theory on
the manifold $(S^1 \cup S^1) \times \R$ and which passes only through 
theories with ground state energy $-\pi/12L$.  In such an adiabatic
process, we expect that only a small amount of energy would be
created.  As opposed to the results of \cite{AD}, this would then
argue that topology change might be possible and that the amount of
energy produced is small so long as the change proceeds `slowly.'

On the other hand, the construction of a
slowly varying gated theory is far from straightforward.  In particular,
allowing the matrix $M_+$ to be time dependent in any way
turns out to destroy conservation of the
Klein-Gordon flux.  As a result, it is not
clear whether our `continuous path through the space of gated theories'
is in fact continuous in any physically meaningful sense.  Allowing
the gates to mix upward and downward moving flux does not help
matters.  It is possible that a frequency dependent gate might be
better behaved but the author, at least, has had no success in this
direction.  

Note Added:  Much the same model of topology change is discussed in 
\cite{Bal} in the context of single particle quantum mechanics.  This
reference, however, also considers the matrix that represents the
boundary conditions to be a dynamical quantum object in its own right.
It would be interesting to see what effect such a treatment would have
on the difficulties encountered here for the field theoretic case.

\acknowledgments
This work was supported by NSF grant
PHY95-97965.
The author would like to thank Gary Horowitz and Anupam Singh for 
a number of related discussions.

\end{document}